\documentclass[3p,times,preprint,authoryear]{elsarticle}

\usepackage{epsfig}

\usepackage{amssymb}

\usepackage[a4paper=true,
 breaklinks=true,%
 colorlinks=true,%
 pdfauthor={Stratta et al.},%
 pdftitle={THESEUS: a key space mission concept for Multi-Messenger Astrophysics}%
 ]{hyperref}

\journal{Advances in Space Research}

\usepackage{cite}

\begin{document}

\begin{frontmatter}



\title{THESEUS: a key space mission {\bf concept} for Multi-Messenger Astrophysics}


\author[1]{G. Stratta}
\author[2,3]{R. Ciolfi}
\author[4]{L. Amati}
\author[5]{E. Bozzo}
\author[6]{G. Ghirlanda}
\author[4]{E. Maiorano}
\author[4]{L. Nicastro}
\author[4]{A. Rossi}
\author[7]{S. Vinciguerra}
\author[8,4]{F. Frontera}
\author[9]{D. G\"{o}tz}
\author[8]{C. Guidorzi}
\author[10]{P. O'Brien}
\author[10]{J. P. Osborne}
\author[11]{N. Tanvir}
\author[12,13]{M. Branchesi}
\author[14]{E. Brocato}
\author[15]{M. G. Dainotti}
\author[16]{M. De Pasquale}
\author[17]{A. Grado}
\author[18]{J. Greiner}
\author[19,20]{F. Longo}
\author[21,22]{U. Maio}
\author[23]{D. Mereghetti}
\author[23,24]{R. Mignani}
\author[25]{S. Piranomonte}
\author[26,27]{L. Rezzolla}
\author[23]{R. Salvaterra}
\author[10]{R. Starling}
\author[10]{R. Willingale}
\author[28]{M. B\"{o}er}
\author[4]{A. Bulgarelli}
\author[29]{J. Caruana}
\author[30]{S. Colafrancesco}
\author[31]{M. Colpi}
\author[6]{S. Covino}
\author[6]{P. D'Avanzo}
\author[32,25]{V. D'Elia}
\author[33]{A. Drago}
\author[4]{F. Fuschino}
\author[34,35]{B. Gendre} 
\author[36,37]{R. Hudec}
\author[38,39]{P. Jonker}
\author[4]{C. Labanti}
\author[40]{D. Malesani}
\author[41]{C. G. Mundell}
\author[4]{E. Palazzi}
\author[42]{B. Patricelli}
\author[42]{M. Razzano}
\author[4]{R. Campana}
\author[8]{P. Rosati}
\author[48]{T. Rodic}
\author[44,45]{D. Sz\'ecsi}
\author[42]{A. Stamerra}
\author[35]{M. van Putten}
\author[47,6]{S. Vergani}
\author[48]{B. Zhang}
\author[49]{M. Bernardini}

\address[1]{Urbino University, via S. Chiara 27, 60129, Urbino (PU, Italy)}
\address[2]{INAF, Osservatorio Astronomico di Padova, Vicolo dell' Osservatorio 5, I-35122 Padova, Italy}
\address[3]{INFN-TIFPA, Trento Institute for Fundamental Physics and Applications, via Sommarive 14, I-38123 Trento, Italy}
\address[4]{INAF-IASF Bologna, via P. Gobetti, 101. I-40129 Bologna, Italy}
\address[5]{Department of Astronomy, University of Geneva, ch. d'\'Ecogia 16, CH-1290 Versoix, Switzerland}
\address[6]{INAF - Osservatorio astronomico di Brera, Via E. Bianchi 46, Merate (LC), I-23807, Italy}
\address[7]{Institute of Gravitational Wave Astronomy \& School of Physics and Astronomy, University of Birmingham, Birmingham, B15 2TT, United Kingdom}
\address[8]{Department of Physics and Earth Sciences, University of Ferrara, Via Saragat 1, I-44122 Ferrara, Italy}
\address[9]{IRFU/D\'epartement d'Astrophysique,  CEA, Universit\'e  Paris-Saclay, F-91191, Gif-sur-Yvette,  France}
\address[10]{Department of Physics and Astronomy, University of Leicester, Leicester LE1 7RH, United Kingdom}
\address[11]{University of Leicester, Department of Physics and Astronomy and Leicester Institute of Space \& Earth Observation, University Road, Leicester, LE1 7RH, United Kingdom}
\address[12]{Università degli Studi di Urbino Carlo Bo, via A. Saffi 2, 61029, Urbino}
\address[13]{INFN, Sezione di Firenze, via G. Sansone 1, 50019, Sesto Fiorentino, Italy}
\address[14]{INAF - Astronomico di Teramo, Mentore Maggini s.n.c., 64100 Teramo, Italy}
\address[15]{Department of Physics \& Astronomy, Stanford University, Via Pueblo Mall 382, Stanford CA, 94305-4060, USA}
\address[16]{Department of Astronomy and Space Sciences, Istanbul University, Beyazit, 34119, Istanbul, Turkey}
\address[17]{INAF - Capodimonte Astronomical observatory Naples, Via Moiariello 16 I-80131, Naples, Italy}
\address[18]{Max Planck Institute for Astrophysics, Karl-Schwarzschild-Str. 1, 85741 Garching, Germany}
\address[19]{Department of Physics, University of Trieste, via Valerio 2, Trieste, Italy} 
\address[20]{INFN Trieste, via Valerio 2, Trieste, Italy}
\address[21]{Leibniz Institut for Astrophysics, An der Sternwarte 16, 14482 Potsdam, Germany}
\address[22]{INAF-Osservatorio Astronomico di Trieste, via G.~Tiepolo 11, 34131 Trieste, Italy}
\address[23]{INAF - Istituto di Astrofisica Spaziale e Fisica Cosmica Milano, via E. Bassini 15, 20133, Milano, Italy} 
\address[24]{Janusz Gil Institute of Astronomy, University of Zielona G\'ora, Lubuska 2, 65-265, Zielona G\'ora, Poland}
\address[25]{INAF-Osservatorio Astronomico di Roma, Via Frascati 33, I-00040 Monte Porzio Catone, Italy}
\address[26]{Institut f{\"u}r Theoretische Physik, Johann Wolfgang Goethe-Universit{\"a}t, Max-von-Laue-Stra{\ss}e 1, 60438 Frankfurt, Germany} 
\address[27]{Frankfurt Institute for Advanced Studies, Ruth-Moufang-Stra{\ss}e 1, 60438 Frankfurt, Germany}
\address[28]{ARTEMIS, CNRS UMR 5270, Universit\'{e} C\^{o}te d'Azur, Observatoire de la C\^{o}te d'Azur, boulevard de l'Observatoire, CS 34229, F-06304 Nice Cedex 04, France}
\address[29]{Department of Physics \& Institute of Space Sciences \& Astronomy, University of Malta, Msida MSD 2080, Malta}
\address[30]{School of Physics, University of Witwatersrand, Private Bag 3, Wits-2050, Johannesburg, South Africa}
\address[31]{Dipartimento di Fisica G. Occhialini, Università degli Studi di Milano Bicocca \& INFN, Sezione di Milano-Bicocca, Piazza della Scienza 3, 20126 Milano, Italy}
\address[32]{Space Science Data Center (SSDC), Agenzia Spaziale Italiana, via del Politecnico, s.n.c., I-00133, Roma, Italy}
\address[33]{INFN, Via Enrico Fermi 40, Frascati, Italy}
\address[34]{University of the Virgin Islands, 2 John Brewer's Bay, 00802 St Thomas, US Virgin Islands}
\address[35]{Etelman Observatory, Bonne Resolution, St Thomas, US Virgin Islands}
\address[36]{Czech Technical University, Faculty of Electrical Engineering, Prague 16627, Czech Republic}
\address[37]{Kazan Federal University, Kazan 420008, Russian Federations}
\address[38]{SRON, Netherlands Institute for Space Research, Sorbonnelaan 2, NL-3584~CA Utrecht, The Netherlands} 
\address[39]{Department of Astrophysics/IMAPP, Radboud University, P.O.~Box 9010, NL-6500 GL Nijmegen, The Netherlands}
\address[40]{Dark Cosmology Centre, Niels Bohr Institute, University of Copenhagen, Juliane Maries Vej 30, DK-2100 Copenhagen, Denmark}
\address[41]{Department of Physics, University of Bath, Claverton Down, Bath BA2 7AY, United Kingdom}
\address[42]{Scuola Normale Superiore, Piazza dei Cavalieri 7, I-56126 Pisa, Italy}
\address[43]{SPACE-SI, Slovenian Centre of Excellence for Space Sciences and Technologies, Ljubljana, Slovenia}
\address[44]{Astronomical Institute of the Czech Academy of Sciences, Fri\v{c}ova 298, 25165 Ond\v{r}ejov, Czech Republic}
\address[45]{School of Physics and Astronomy and Institute of Gravitational Wave Astronomy, University of Birmingham, Edgbaston, Birmingham B15 2TT, United Kingdom}
\address[46]{Sejong University, 98 Gunja-Dong Gwangin-gu, Seoul 143-747, Korea}
\address[47]{GEPI, Observatoire de Paris, PSL Research University, CNRS, Place Jules Janssen, 92190 Meudon}
\address[48]{Department of Physics and Astronomy, University of Nevada, Las Vegas, NV 89154, USA}
\address[49]{Universit\'e Montpellier 2, Campus Triolet, Place Eugene Bataillon - CC 070, 34095 Montpellier Cedex 5}

\begin{abstract}
The recent discovery of the electromagnetic counterpart of the gravitational wave source GW170817, 
has demonstrated the huge informative power of multi-messenger observations. 
During the next decade the nascent field of multi-messenger astronomy will mature significantly. Around 2030, 
third generation gravitational wave detectors will be roughly ten times more sensitive than the current ones. At the same time, neutrino detectors 
currently upgrading to multi km$^3$ telescopes, will include a 10 km$^3$ facility in the Southern hemisphere that is expected to be operational around 2030.  
In this review, we describe the most promising high frequency gravitational wave and neutrino sources that will be detected in the next two decades. 
In this context, we show the important role of the {\it Transient High Energy Sky and Early Universe Surveyor}  (THESEUS), a mission concept proposed 
to ESA by a large international collaboration in response to the call for the Cosmic Vision Programme M5 missions. 
THESEUS aims at providing a substantial advancement in early Universe science as well as playing a fundamental role in multi--messenger 
and time--domain astrophysics, operating in strong synergy with future gravitational wave and neutrino detectors 
as well as major ground- and space-based telescopes. This review is an extension of the THESEUS 
white paper \citep{Amati2017}, also in light of the discovery of GW170817/GRB170817A that was announced on October 16$^\mathrm{th}$, 2017.
\end{abstract}


\end{frontmatter}

\parindent=0.5 cm

\section{Introduction}

With the first detection in 2015 of gravitational waves (GWs) from black hole binary systems during their  coalescing phase 
\citep{Abbott2016b, Abbott2016a}, a new observational window on the Universe has been opened. Stellar-mass black hole coalescences, 
together with binary neutron star (NS-NS), NS-black hole (BH) mergers, burst sources as core-collapsing massive stars and 
possibly NS instability episodes, are among the main targets of ground-based GW detectors\footnote{An ensemble of Michelson-type 
interferometers sensitive to the high frequency range, from few Hz to few thousand Hz}.  
Some of these sources are also expected to produce neutrinos and electromagnetic (EM) signals over the entire spectrum, 
from radio to gamma-rays.  

These expectations were astonishingly satisfied for the first time on August 17$^\mathrm{th}$, 2017, when a GW signal 
consistent with a binary neutron star merger system \citep{Abbott2017a} was found shortly preceding the short gamma-ray 
burst GRB170817A \citep{Abbott2017b}. The GW170817 90\% confidence sky area obtained with the Advanced LIGO \citep{Harry2010} 
and Advanced Virgo \citep{Acernese2015} network was 
fully contained within the GRB error box. In addition, a ``kilonova" (or ``macronova")  emission (AT2017gfo), theoretically 
predicted from such systems \citep[e.g.][]{Li1998}, has been found within the GW-GRB error-box and positionally consistent 
with NGC4993, a lenticular galaxy at a distance compatible with the GW signal 
\citep{Abbott2017c,Smartt2017,Tanvir2017,Pian2017,Coulter2017}. 

\begin{figure*}[t!]
\centering
\includegraphics[scale=0.52]{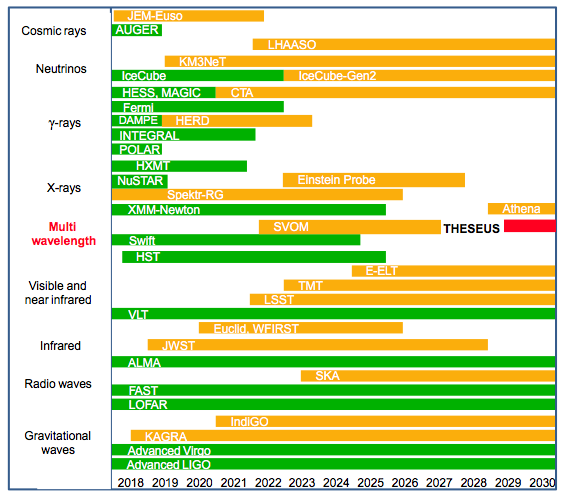}
\caption{THESEUS within the multi-messenger Astrophysics context of 2020-2030. Green and orange labels are for presently 
operating and future planned or under construction instruments (Figure credit: S. Schanne). }
\label{fig:context}   
\end{figure*}

\begin{figure*}
\centering
\includegraphics[scale=0.72]{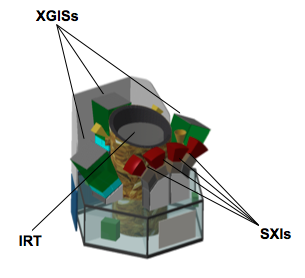}
\caption{THESEUS Satellite Baseline Configuration and Instrument suite accommodation.}
\label{fig:payload}   
\end{figure*}
   
\begin{table*}
\caption{THESEUS instruments }
\label{tab:main}  
\resizebox{\columnwidth}{!}{
\begin{tabular}{ll | cccc | lc}
\hline\noalign{\smallskip}
                          		& SXI         						&  \multicolumn{4}{c}{XGIS}        &  &IRT \\
\noalign{\smallskip}\hline\noalign{\smallskip}

Energy range              	& 0.3-6 keV 						& 		     & 2 -30 keV  			& 30-150 keV & $>150$ keV &	&ZYJH   \\

 &&&&&&& ($0.7-1.8 \mu$m)\\

\hline &&&&&&&\\
Field of View             		&   1 sr    							& Half sens.: &  50$\times$50 deg$^2$  	& 50$\times$50 deg$^2$ &  & imaging & $10'\times10'$ \\
		             		&   	   							& Total:  	    & 64$\times$64 deg$^2$  	& 85$\times$85 deg$^2$& $2\pi$ sr & low res & $10'\times10'$\\
					&   	   							& 	    	     & 					     	& 				    & 		 	& high res & $5'\times5'$\\

\hline &&&&&&&\\
Source location   		&  $<10''$ (best)					&          		&    $5'$ (for $>6 \sigma$ source)   & -&- & &  $<1''$   \\
accuracy 		 		&  $105''$ (worse) 					&	   		 &							   & & & &\\

\hline &&&&&&&\\
Sensitivity 			& erg(ph) cm$^{-2}$ s$^{-1}$  			& & ph cm$^{-2}$ s$^{-1}$   &         		&  		& &  H (AB mag)\\
		 			&  $2\times10^{-8}$(10)  (1s)  			& & 1 (1s)   			   &  0.15 (1s)    & 0.22 (1s) &  imaging & 20.6 (300 s) \\
               				& $2\times10^{-11}$(0.01) (10 ks)		& &  0.02 (1ks)  		   & 0.004 (1ks)    & 0.008 (1ks) &  low res. & 18.5 (300 s) \\
		               		&  								& &     				   &     		& 			&  high res. & 17.5 (1800 s) \\
\noalign{\smallskip}\hline
\end{tabular}
}
\end{table*}

By the end of the twenties, the sky will be routinely monitored by the second-generation GW detector network, 
composed by the two Advanced LIGO (aLIGO) detectors in the US, Advanced Virgo 
in Italy, ILIGO in India \citep[e.g.][]{Abbott2016} and KAGRA in Japan \citep{Somiya2012}. 
Then, around 2030, more sensitive third generation ground-based GW interferometers, such as the Einstein Telescope \citep[ET,  e.g.][]{Punturo2010} 
and LIGO Cosmic Explorer \citep[LIGO-CE, e.g.][]{Abbott2017d}, are planned to be operational and to provide an increase of 
roughly one order of magnitude in sensitivity. In parallel to these advancements, IceCube and KM3nNeT and the advent of 
10 km$^3$ detectors \citep[e.g. IceCube-Gen2,][and references therein]{Aartsen2014} will enable to gain high-statistics 
samples of astrophysical neutrinos.  The 2030 will therefore coincide with a golden era of multi-messenger astrophysics (MMA, Figure 1). 

By that time, the ESA M5 approved missions for space-based astronomy will be launched. 
THESEUS ({\it Transient High Energy Sky and Early Universe Surveyor}\footnote{\url{http://www.isdc.unige.ch/theseus}} \citealt{Amati2017}) is a space mission concept developed by a large International collaboration currently under evaluation by ESA within the selection process 
for next M5 mission of the Cosmic Vision Programme.
If selected, the launch of THESEUS (2029) will provide a very strong contribution to MMA. 
In the following sections, after a short review of the main characteristics \citep[\S \ref{theseus}; see][for a more exhaustive 
description of the mission concept]{Amati2017},  we describe the role of THESEUS is the MMA and 
the most promising GW (\S\ref{gw}) and neutrino (\S\ref{nu}) sources that THESEUS will observe. We also provide the expected joint 
GW+EM detection rates for the most promising GW+EM sources (e.g. NS-NS) taking into account the facilities planned to be operational by the end of the twenties. 

\begin{figure*}[t!]
\centering
\includegraphics[scale=0.5]{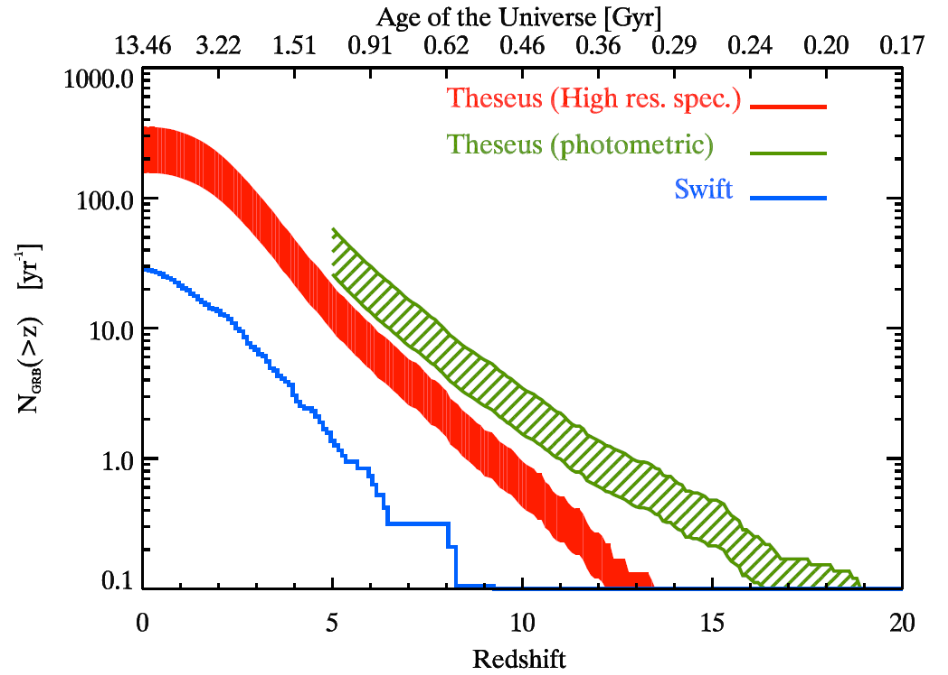}
\caption{The yearly cumulative distribution of GRBs with redshift determination as a function of the redshift 
for Swift and THESEUS \citep{Amati2017}. The THESEUS expected improvement in the detection and identification of GRBs 
at very high redshift
w/r to present situation is impressive (more than 100--150 GRBs at z$>$6 and several tens at z$>$8 in a few years) and
will allow the mission to shade light on main open issues early Universe science (star formation rate evolution,
re--ionisation, pop III stars, metallicity evolution of first galaxies, etc.).}
\label{fig:logn-logz}   
\end{figure*}

\begin{figure*}[t!]
\centering
\includegraphics[scale=0.5]{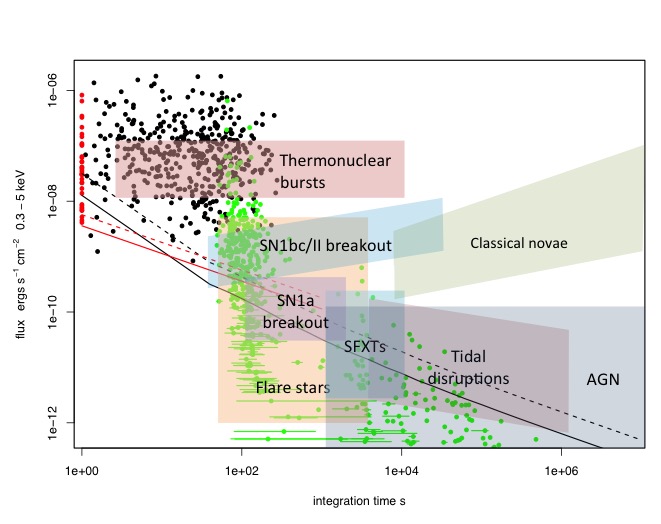}
\caption{Sensitivity of the SXI (black curves) and XGIS (red) vs. integration time \citep{Amati2017}. 
The solid curves assume a source column density of $5\times10^{20}$  cm$^{-2}$ (i.e., well out of 
the Galactic plane and very little intrinsic absorption). The dotted curves assume a source column density 
of $10^{22}$ cm$^{-2}$ (significant intrinsic absorption). The black dots are the peak fluxes for Swift 
BAT GRBs plotted against T90/2 (where T90 is defined as the time interval over which 90\% of the total 
background-subtracted counts are observed, with the interval starting when 5\% of the total counts have 
been observed, \citealp{Koshut1995}). The flux in the soft band 0.3-10 keV was estimated using the T90 
BAT spectral fit including the absorption from the XRT spectral fit. The red dots are those GRBs for which 
T90/2 is less than 1 s. The green dots are the initial fluxes and times since trigger at the start of the Swift 
XRT GRB light-curves. The horizontal lines indicate the duration of the first time bin in the XRT light-curve. 
The various shaded regions illustrate variability and flux regions for different types of transients and variable sources.
}
\label{fig:survey}   
\end{figure*}

\section{The THESEUS Mission}
\label{theseus}

The THESEUS mission aims at exploiting GRBs for investigating the early Universe and at providing a substantial 
advancement in multi-messenger and time-domain astrophysics \citep[see][for a detailed review]{Amati2017}.

The instrumentation foreseen on board, illustrated in Figure \ref{fig:payload}, includes:
\begin{itemize}
\item Soft X-ray Imager (SXI, 0.3-6 keV): a set of 4 lobster-eye telescopes units, covering a total FoV of $\sim$1 sr with source location accuracy $<1$ arcmin;
\item X-Gamma ray Imaging Spectrometer (XGIS, 2 keV-20 MeV): a set of coded-mask cameras using monolithic X-gamma
ray detectors based on bars of Silicon Drift Diodes coupled
with CsI crystal scintillator, granting an unprecedentedly broad energy band, a FoV up to $\sim$4sr, a source location 
accuracy of $\sim$5 arcmin, and an energy resolution of $\sim$200--300 eV in 2-30 keV; 
\item InfraRed Telescope (IRT, 0.7-1.8 $\mu$m): a 0.7 m class IR telescope with $10\times10$ arcmin FoV, for fast 
response, with both imaging and spectroscopy capabilities.
\end{itemize}

The main characteristics and sensitivities of these instruments are summarised in Table~\ref{tab:main}. 
The mission profile includes fast slewing capability, allowing to point the IRT to the position of GRBs and of 
other transient sources detected and localised by the SXI and/or the XGIS.   
Fast slewing observations will enable the possibility of promptly transmitting to ground trigger time, position, and redshift of these events 
(as evaluated on--board by means of IRT photometry and spectroscopy), thus enabling quick follow--up with large ground-- and space--based multi--wavelength 
observatories. As shown in Figure \ref{fig:logn-logz} and detailed in \citet{Amati2017}, 
this unique combination of scientific instruments and mission
profile will allow THESEUS to make a giant leap in the use of Gamma--Ray Bursts for shading light on the main
open questions on the early Universe (star formation rate evolution up to the end of "dark ages", cosmic re--ionisation,
metallicity evolution of the early galaxies, pop III stars, ...). In the next section we describe how fast slewing capabilities of IRT 
will be of great relevance also for multi-messenger observational campaigns.   

THESEUS will be also used as a flexible infrared observatory complementary to other facilities, 
as it is the case for the Swift mission in X-rays and UV (see \citealt{Amati2017}).  
The SXI can localise to better than an arcminute, and sometimes tens of arcsec depending on the source  count rate, thus significantly better than the XGIS. 
The large grasp\footnote{Grasp is an appropriate measure for large area, medium-deep surveys when monitoring on timescales of the order of seconds to days, over large sky areas.} {\bf (i.e. FoV$\times$Effective Area)} of the SXI, joined with the broad energy band, large effective area and few arcmin source location accuracy 
of the XGIS, will enable the discovery and study of a wealth of transient sources, both Galactic and extra-galactic 
(Fig. \ref{fig:survey}) many of which are expected to be also neutrino and/or GW sources (e.g. GRB, SGR, CCSNe, AGN). 

THESEUS will observe in synergy with several telescopes operating at different wavelengths, as illustrated in Figure \ref{fig:context}, 
among which it is worth mentioning: 1) the space-based telescopes James Webb Space Telescope (JWST), ATHENA and WFIRST; 2) the ground-based telescopes with large FoV like zPTF and LSST; 3) the 30-m class  telescopes GMT, TMT and ELT; 4) the Square Kilometer Array (SKA) in the radio; 5) the very high-energy (GeV-TeV) Cherenkov Telescope Array (CTA). We note that the main differences between THESEUS and the other large X-ray telescope facility operational around 2030, ATHENA, are the much larger field of views of the X-ray and gamma-ray detectors on board THESEUS that will make it a ``surveyor" instrument, and the presence of an infrared telescope with both imaging and spectroscopic capabilities.

\begin{figure*}[t!]
\centering
\includegraphics[scale=0.5]{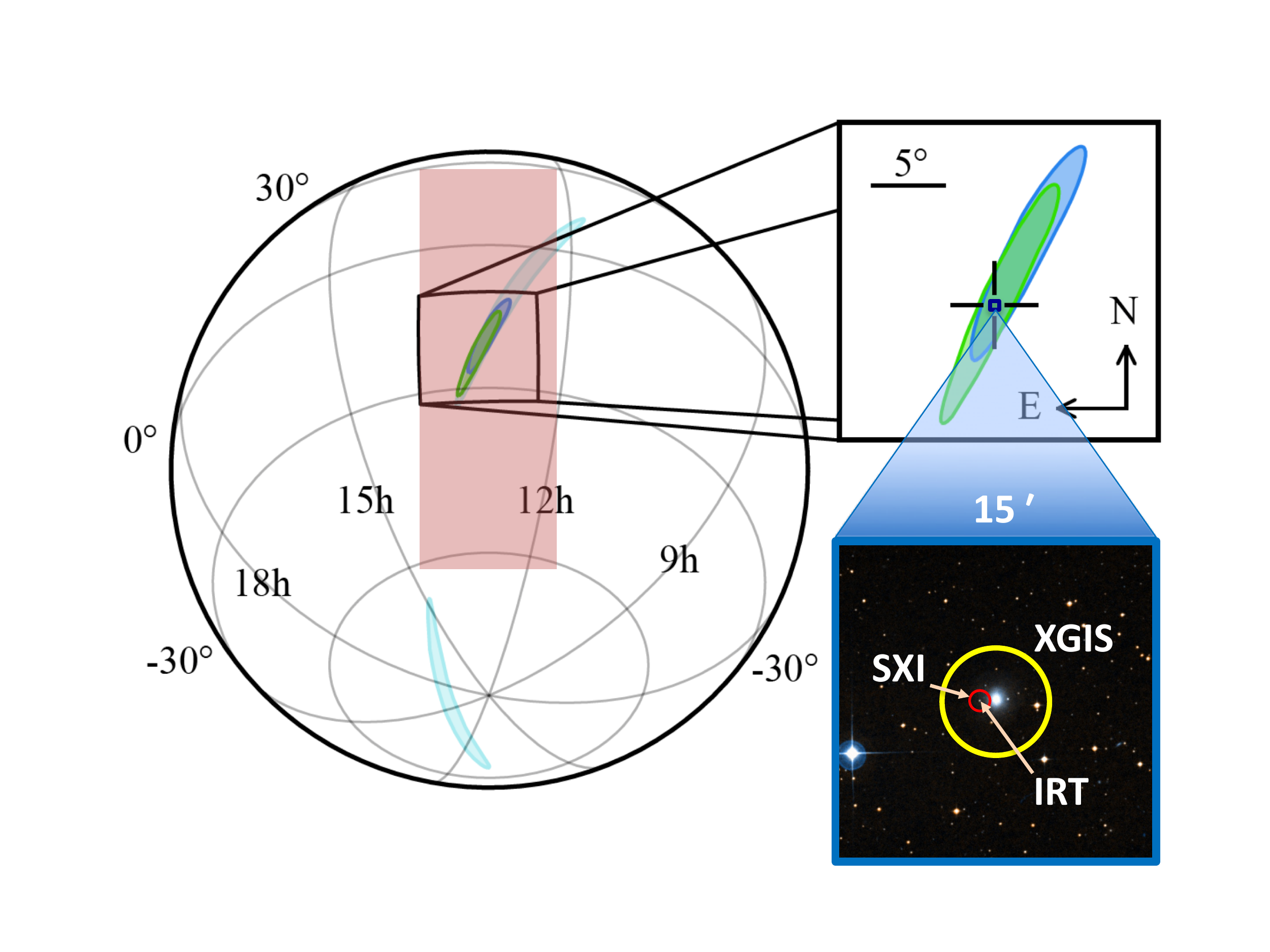}
\caption{
The plot shows the THESEUS/SXI field of view ($\sim 110\times30$~deg$^2$, pink rectangle) superimposed on the probability skymap of 
GW 170817 obtained with the two Advanced LIGO only (cyan) and with the addition of Advanced Virgo (green) \citep{Abbott2017a}. 
THESEUS not only will cover a large fraction of the skymap (even those obtained with only two GW-detectors, e.g. cyan area), but will 
also localise the counterpart with uncertainty of the order of 5 arcmin with the XGIS and to less than 1 arcmin with SXI. 
{\it The THESEUS location accuracy of GW events produced by NS-NS mergers can be as good as 1~arcsec in case of detection 
of the kilonova emission by the IRT}. By the end of the 2020s, if ET will be a single detector, almost no directional 
information will be
available for GW sources ($>1000$ deg$^2$ for BNS at $z>0.3$, \citealt[e.g.][]{Zhao2017}), and a GRB-localising 
satellite will be essential to discover EM counterparts.}
\label{fig:SXI_FOV_GW170817}   
\end{figure*}

\section{The role of THESEUS in the Multi-Messenger Astronomy}  

The detection of EM counterparts of GW and neutrino signals will enable a multitude of science 
programmes \citep[see, e.g.,][]{Bloom2009, Phinney2009} by allowing for parameter constraints that the GW or neutrino observations alone cannot fully provide. 
GW detectors have relatively poor sky localisation capabilities, mainly based on triangulation methods, that on average will not be better than 
few dozens of square degrees \citep{Abbott2016}. 
For GW sources at distances larger than the horizon of second-generation detectors (200 Mpc), therefore accessible only by the third-generation 
ones around 2030  (e.g. Einstein Telescope and Cosmic Explorer, \citealt{Punturo2010,Abbott2017b}), sky localisation may even worsen 
if the new generation network will be composed by only one or two detectors, with possible values of the order of few hundred 
square degrees or more \citep[e.g.][]{Zhao2017}. 
Neutrino detectors can localise to an accuracy of better than a few squares degrees 
\citep[see, e.g.,][and references therein]{Santander2016}. 
In order to maximise the science return of the multi-messenger investigation it is essential to have 
a facility that (i) can detect and disseminate an EM signal independently to the GW/neutrino event and (ii) can rapidly search with good sensitivity in the large error boxes provided by the GW and neutrino facilities.  

These combined requirements are uniquely fulfilled by THESEUS. 
The hard XGIS and/or SXI will trigger and localise transient sources within the uncertain GW and/or neutrino error boxes. 
In particular, a very large fraction of the error boxes of poorly localised GW sources can be covered with SXI FoV 
within one orbit due to the large grasp of the instrument (see \citealt{Amati2017} and Fig. 5).  In response to an SXI/XGIS 
trigger, if an optical counterpart is present, arcsecond localisations can be obtained with IRT and disseminated within minutes to the astronomical community, 
thus enabling large ground-based telescopes to observe and deeply characterise the nature of the GW/neutrino source. 
Figure \ref{fig:survey} clearly show how the wide FoV of THESEUS will guarantee autonomous triggers of a large number of transient X-ray and 
gamma-ray sources. This will enable {\it independent} trigger of the EM counterpart of several GW/neutrino sources, as 
it was the case for GRB170817A triggered by Fermi/GBM. However, with respect to the Fermi/GBM, THESEUS will provide also accurate localisation,  as sketched in Figure \ref{fig:SXI_FOV_GW170817}. 
In response to THESEUS triggers, GW and neutrino archival data analysis will enable to search for simultaneous events 
at the time of the trigger (e.g., due to GRBs or supernovae), since these type of detectors record all their data almost continuously. 
This strategy has been already pursued by the LIGO-Virgo collaboration for a number of GRBs \citep[e.g.][]{Abbott2005,Abbott2008,Abbott2017e}. 

\begin{figure*}
\centering
\includegraphics[scale=0.70]{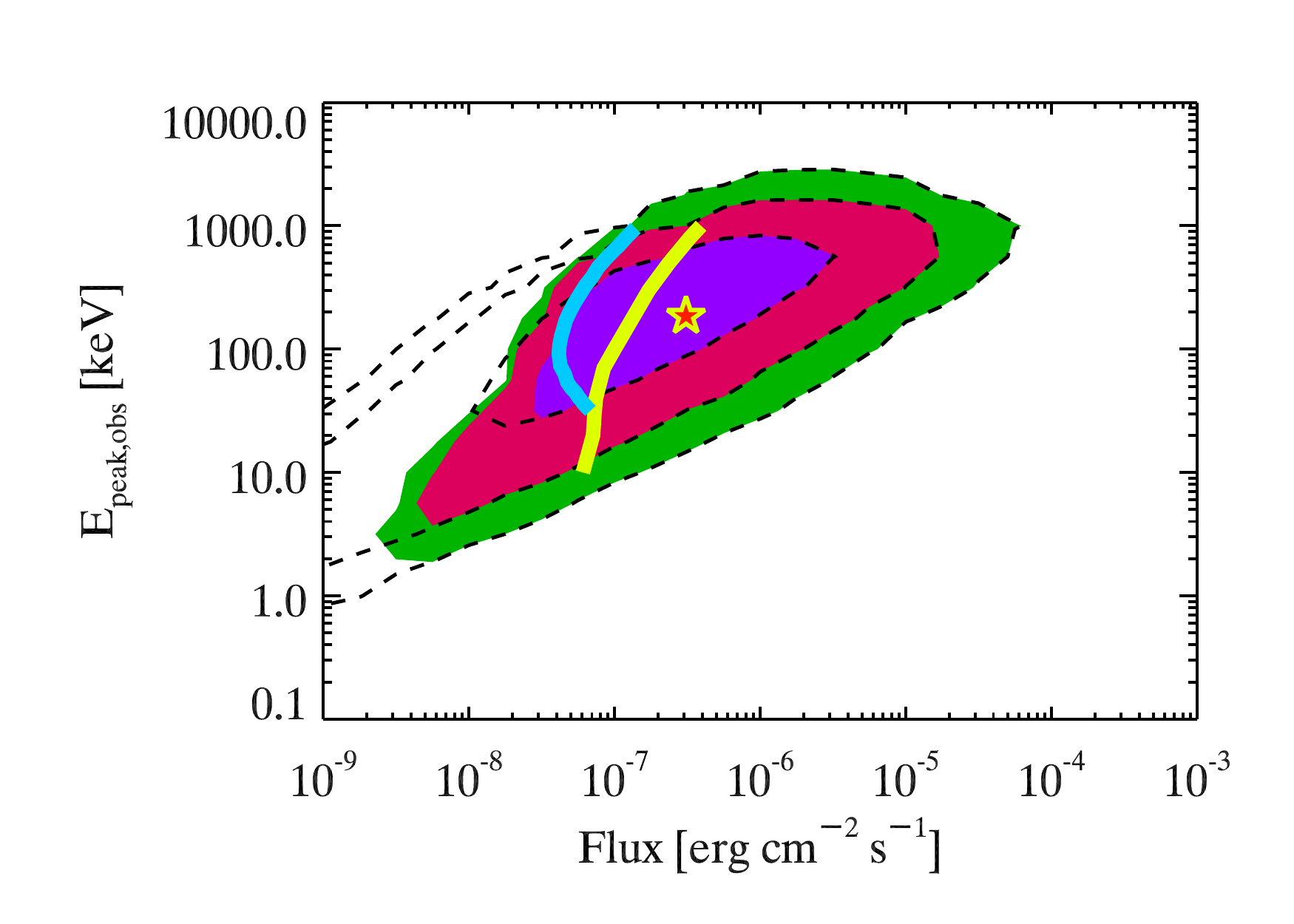}
\caption{Density contours (dashed lines) corresponding to  1, 2, 3 $\sigma$ levels of the synthetic population of Short GRBs 
(from \citealt{Ghirlanda2016}). Shaded coloured regions show the density contours of the population detectable by THESEUS. 
The yellow and cyan lines show the trigger threshold of Fermi/GBM and GCRO/BATSE (from \citealt{Nava2011}). 
The flux is integrated over the 10-1000 keV energy range. The star symbol shows the short GRB170817A \citep{Goldstein2017}.}
\label{fig:capabilities}   
\end{figure*} 

\begin{figure*}
\centering
\includegraphics[scale=0.70]{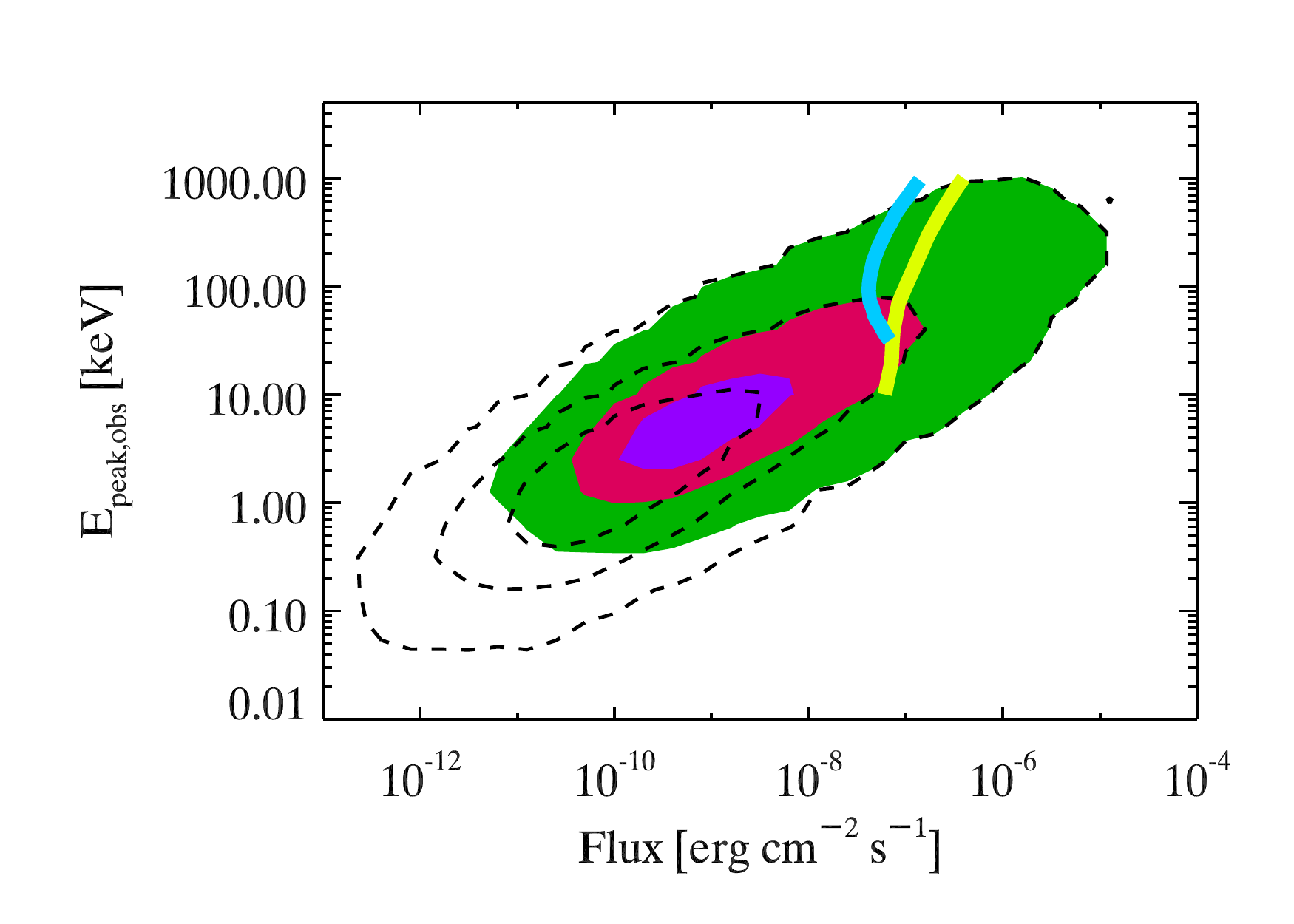}
\caption{Density contours (dashed lines) corresponding to  1, 2, 3 $\sigma$ levels of the synthetic population of Long GRBs 
(from \citealt{Ghirlanda2015c}). Shaded coloured regions show the density contours of the population detectable by THESEUS. 
The yellow and cyan lines show the trigger threshold of Fermi/GBM and GCRO/BATSE (from \citealt{Nava2011}).  
The flux is integrated over the 10-1000 keV energy range. As can be seen, THESEUS will carry onboard the 
ideal instruments suite for detecting all classes of GRBs (classical long GRBs, short/hard GRBs, sub--energetic GRBs,
and very high-redshift GRBs, which, in this plane,  populate the region of weak/soft events), providing a redshift
estimate for most of them \citep{Amati2017}.}
\label{fig:capabilities_long}   
\end{figure*} 

\begin{figure*}[t!]
\centering
\includegraphics[scale=0.35]{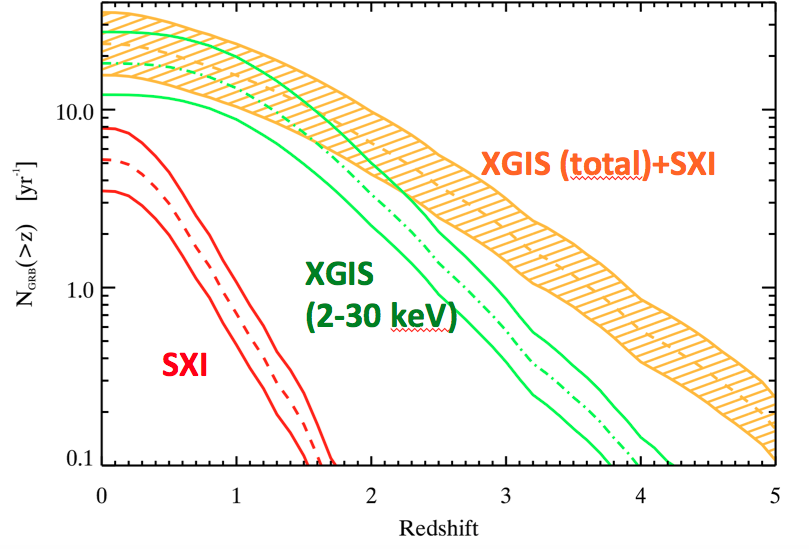}
\caption{Cumulative distribution of the rate of  short GRBs as a function of redshift  that Theseus will detect (yellow stripe 
filled region). The fraction of the population that will be detected by the soft coded mask instruments of XGIS 
(2-30 keV) is shown by the green stripe. The cumulative distribution of the fewer short GRBs also detected by SXI 
is shown by the red stripe. The vertical width of the stripes account for the uncertainties of the model parameters 
of the short GRB population adopted \citep{Ghirlanda2016}.}
\label{fig:rate}   
\end{figure*}

\begin{figure*}[t!]
\centering
\includegraphics[scale=0.55]{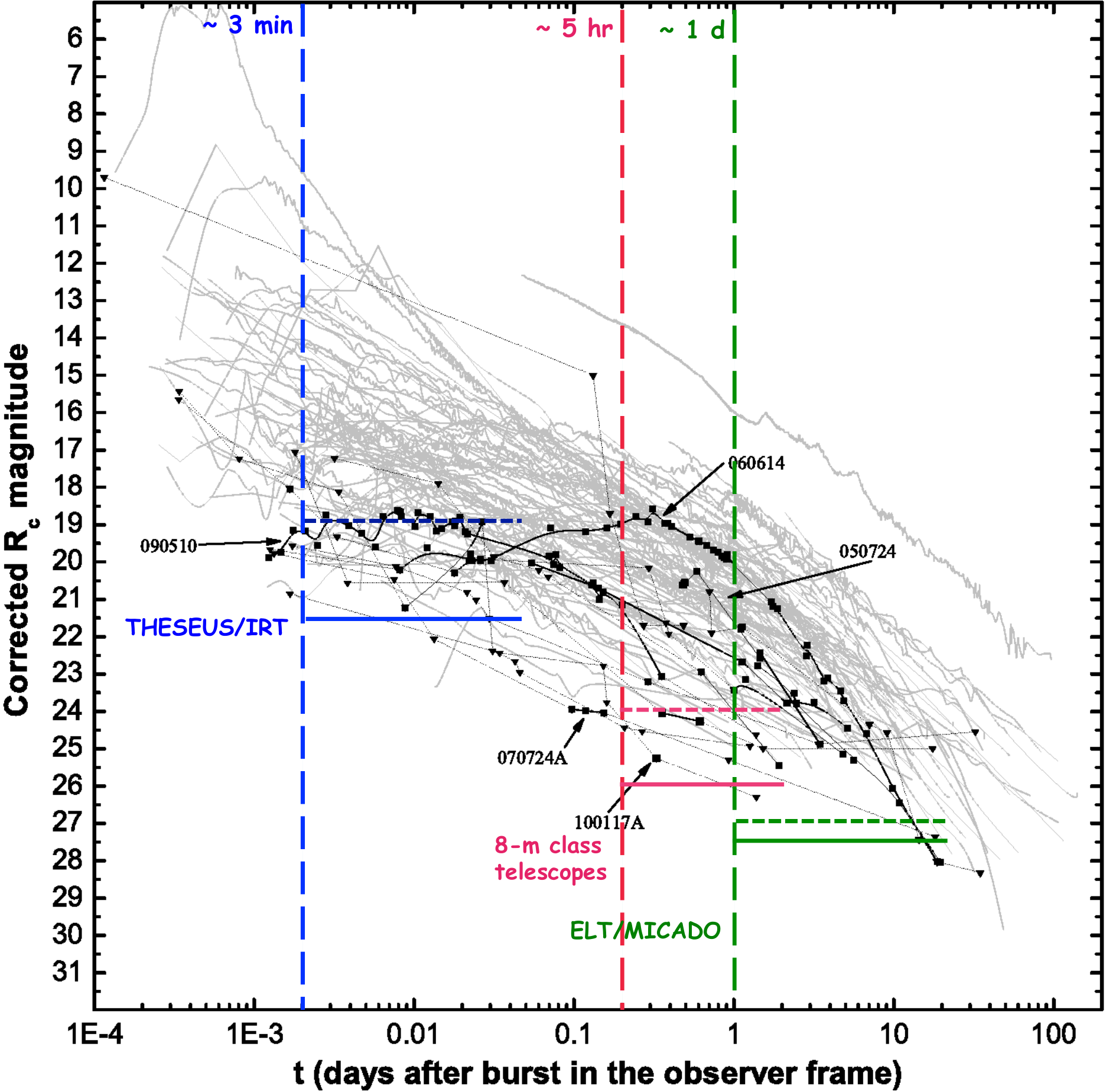}
\caption{R-band light curves of long (grey lines) and short (black dots) GRBs (adapted from \citealt{Kann2011}). 
The limiting magnitudes achievable with THESEUS/IRT with 300 s of exposure (blue lines), 8-m class telescopes (red lines) 
and ELT/MICADO (green lines) are also shown. Dashed horizontal line for the spectroscopy, solid line for the imaging. 
The magnitudes are rescaled from H-band to R-band assuming achromatic behaviour and a spectral index beta=0.7. 
A tentative observation strategy could consist of the following steps: first starting the follow up with IRT, 
then activating the observations with 8-m class telescopes after few hours, finally, according to the brightness 
of the afterglow and thanks to the very high sensitivity of ELT, performing late observations for weeks.}
\label{fig:afterglow}   
\end{figure*}

Several multi-wavelength and multi-messenger sources are among the main targets of THESEUS, as for example GRBs, 
flaring magnetars, core-collapse supernovae (CCSNe) and AGNs \citealt{Amati2017}. 
The recent association of GW170817 with the short GRB 170817 makes the short GRB detection capabilities of crucial 
relevance for MMA in an epoch where almost all short GRBs will be accompanied by a 
GW signal detected by the third-generation interferometers (e.g. ET or LIGO-CE).  
Figure \ref{fig:capabilities} shows the density contours of the population of short GRBs (dashed contours) in the peak energy - peak flux 
plane \citep{Ghirlanda2016}. Due to their harder spectrum, short GRBs are better triggered by THESEUS/XGIS than SXI. 
Compared to the detection thresholds of BATSE and Fermi/GBM, THESEUS will slightly extend the detected population towards lower fluxes and softer peak energies. The plot shows how THESEUS will be able to fully access to events similar to GRB170817 and explore their nature.  
Although the XGIS sensitivity threshold improves over GBM, its smaller (by a factor of 2) FoV compensates this gain reaching 
a detection rate of short GRBs which is comparable to that of GBM, that is about 15-35 per year (Fig. \ref{rate}).  What makes THESEUS XGIS unique, with respect to GBM, is the possibility 
to locate, thanks to the soft (2 keV-30 keV) coded mask detectors of the XGIS, most of the detected short GRBs with an expected accuracy 
of 5 arcmin (to be compared with the average $>$ few degrees of GBM GRBs). 

Figure \ref{fig:capabilities_long} shows the density contours (dashed lines) of synthetic population of long GRBs 
\citep{Ghirlanda2015c} in the observer--frame plane representing the peak energy $E_{\rm peak}$ versus the (10-1000 keV) 
peak flux.  Given the association of long GRB to CCSNe and the expected GW radiation as well as neutrino emission of these events (see Section 4.3), long GRBs   detection capabilities are also relevant for multi-messenger joint observational campaigns.  
The plot shows how THESEUS will access a region of the $E_{\rm peak}$-Peak flux plane totally unexplored by past and current instruments. 
A large fraction of its population will be constituted by soft low flux events. Among these there will be (i) low redshift/low luminosity events 
\citep[with a $E_{\rm peak}$ due to the correlation between these two observables;][]{Yonetoku2004} which 
are particularly relevant for simultaneous GW and/or neutrino detections  
and (ii) long GRBs at high redshifts which, used as beacons, will allow us to explore the high redshift Universe and its evolution and for which  the redshift will be measured on board using by THESEUS/IRT (Fig. \ref{fig:afterglow}).
A good fraction of optical/NIR afterglows detected by space- and ground-based observatories would be immediately visible by IRT. In particular long GRBs afterglows would be detectable both by imaging and low-resolution spectroscopy (LRS), making IRT capable to measure redshift for all $z>5$ GRBs, after constraining the Lyman drop-out.  Short GRBs will be harder to observe spectroscopically, however, thanks to the THESEUS/IRT arcsecond localisation of the afterglows, ground and space based telescope will be able will be able to follow-up
the IRT afterglows hours and days after the trigger (Fig. \ref{fig:afterglow}).

Besides the expected collimated GRB ``prompt" emission, softer X-ray emission is also expected from the side and/or afterglow emission of the GRB jet, with a much lower degree of collimation (see \S 4.2). For short GRB sources and in particular NS-NS mergers, an additional nearly isotropic soft X-ray emission is possibly expected when the merger remnant is a long-lived NS or magnetar, where the corresponding transient is powered by the spindown of the latter (see S \ref{isotropicemission}).

\subsection{Science return from joint GW+EM detections with THESEUS}
\begin{figure*}[t]
\centering
\includegraphics[scale=0.50]{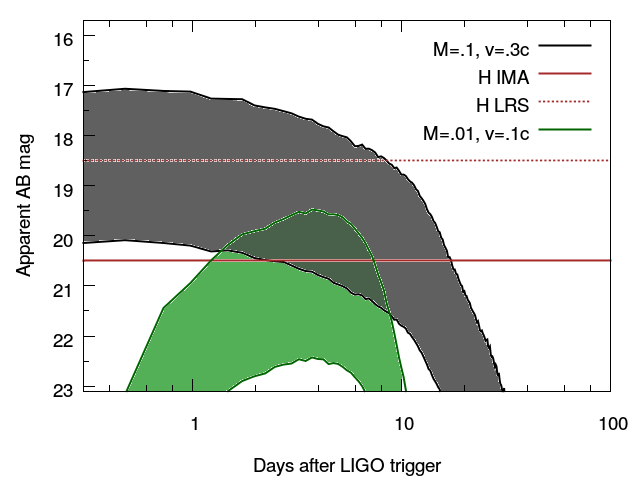}
\caption{
Theoretical $H$-band lightcurves of kilonova based on models \citep[from][]{Barnes2016}. The lightcurves are in observer 
frame for a source between 50 and 200 Mpc. Grey model is for the most optimistic case of a kilonova with $0.1\, M_\odot$ 
ejected mass with speed of $0.3\,c$. Green model is for a weaker emission, corresponding to $0.01\, M_\odot$ ejected mass 
with speed of $0.1\,c$. The continuous and dashed red lines indicate the THESEUS/IRT limiting H magnitudes for imaging 
and prism spectroscopy, respectively, with 300 s of exposure \citep[see][]{Amati2017}. }
\label{fig:kilonova}   
\end{figure*}

Each individual joint observation of an EM source and its GW and/or neutrino counterpart, provides an enormous science return.  
To mention just few examples in the case of compact binary coalescences:
{\it i)} the determination of the GW polarisation ratio would constrain the binary orbit inclination and hence, when combined 
with an EM signal, the jet geometry and source energetics; 
{\it ii)} a better understanding of the NS equation of state can follow from combined GW and X-ray emission signals 
(see \S \ref{isotropicemission}) 
\citep[see, e.g.,][]{Bauswein2012,Takami2014,Lasky2014,Ciolfi2015a,Ciolfi2015b,Messenger2015,Rezzolla2016,Drago2016};
{\it iii)} an estimate of the amount of matter expelled during a NS-NS or a NS-BH merger \citep[e.g.][and references therein]{Fernandez2016}; 
{\it iv)} tracing the history of heavy-metal enrichment of the Universe by promptly follow-up the kilonova/macronova IR emission, as shown in Figure~\ref{fig:kilonova};  
{\it v)} redshift measurements of a large sample of short GRBs combined with the absolute source luminosity distance 
provided by the CBC-GW signals can deliver precise measurements of the Hubble constant \citep{Schutz1986}, 
helping to break the degeneracies in determining other cosmological parameters via CMB, SNIa and BAO surveys \citep[see, e.g.,][]{Dalal2006}. 
The last point on the Hubble constant measure is of particular relevance for THESEUS especially during the third-generation 
GW detector era, when, as we will show in section 4, almost all gamma-ray detected short GRBs will have a GW counterpart. 
The onboard IRT will enable to measure the redshift for all those events far away (mid to high redshifts) or hosted in faint galaxies for which 
there is no adequate galaxy catalog. When not taking spectra, it will transmit precise localisation to large size telescopes to get redshift measurements.
A first attempt of Hubble constant measurement has been explored with GW170817 for which the recession velocity $v_r$ 
of the optical transient AT2017gfo host galaxy NGC4993, was combined with the luminosity distance $D_L$ measured directly from the waveform of GW170817. 
For small distances, as in this case ($\sim40$ Mpc), the Hubble constant depends only on these two variables as $H_0=v_r/D_L$. 
Despite the large uncertainties on this first measurement, the results are very encouraging \citep{Abbott2017f}. The value obtained, 
$H_0 =70.0^{+12.0}_{-8.0}$ km s$^{-1}$  Mpc$^{-1}$, lies in between the measurements obtained from SNIa from SHoES 
($73.24\pm1.74$ km s$^{-1}$ Mpc$^{-1}$, \citep{Riess2016} and CMB from Planck 
($67.74\pm0.46$ km s$^{-1}$ Mpc$^{-1}$, \citep{Planck2016a}. 
Furthermore, combining the observing angle vs.~GW amplitude degeneracy measured by LIGO-Virgo interferometers with 
independent information on the observing angle derived from modelling the associated broadband  afterglow, 
a further reduction by $\sim5$\% on the uncertainty interval of $H_0$ could be obtained \citep{Guidorzi2017}.

\begin{figure*}[t]
\centering
\includegraphics[scale=0.75]{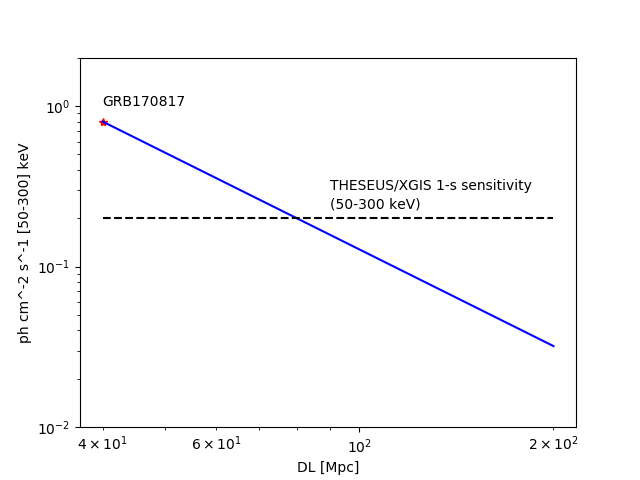}
\caption{
The Fermi/GBM peak photon flux of the short GW/GRB 170817 (red star, \citealt{Goldstein2017}) rescaled with the distance (blue line) and compared with THESEUS/XGIS 1-s sensitivity in the 50-300 keV energy range. Off-axis short GRB similar to GRB 170817 could had been detected with THESEUS/XGIS up to $\sim 70-80$ Mpc.}
\label{fig:sens}   
\end{figure*}

\subsection{THESEUS and GRB170817A}

In this section we explore THESEUS capabilities in the detection and characterisation of the short GRB170817A associated 
with GW170817 \citep{Abbott2017a,Goldstein2017,Savchenko2017}.   
The two events were found consistent with being originated from a common source with high confidence ($\sim10^{-7}$ 
probability of being independent, \citealt{Abbott2017a}).
In addition, a bright optical transient was observed in NGC4993 \citep{Smartt2017,Tanvir2017,Pian2017,Abbott2017b,Coulter2017} at $\sim40$ Mpc. 
This was by far the closest short GRB yet observed.  
The gamma-ray peak photon flux ($3.7\pm0.9$ ph cm$^{-2}$ s$^{-1}$ in the 10-1000 keV band, Goldstein et al. 2017) 
implies an extremely low isotropic luminosity short GRB if compared with typical values ($\sim 10^{51} - 10^{53}$ erg s$^{-1}$, 
see e.g. \citealt{Ghirlanda2015a}), with $1.7 \times 10^{47}$ erg s$^{-1}$ (e.g. Zhang et al. 2017). 
Since various indications point at a binary merger seen with viewing angle $\sim20-40$ deg away from the normal direction to the 
orbital plane, a possible explanation of the low luminosity is that the event was a short GRB with a structured jet observed 
off-axis \citep[e.g.][]{Troja2017,Alexander2017,Margutti2017,Haggard2017,Hallinan2017,Lazzati2017c}. 
Within the latter scenario, GRB170817A suggests an extension of the observed short GRB population to include a larger 
fraction of dimmer events \citep[e.g.][]{Burgess2017}, which can enhance the coincident short GRB/GW detection rate up to relatively small distances (i.e. $<$100 Mpc) with sensibile instruments such as THESEUS.
The late-time X-ray and radio counterparts detected respectively $\sim$9 and $\sim$16 days after merger \citep{Troja2017,Margutti2017,Alexander2017, 
Hallinan2017,Haggard2017} 
and the additional X-ray and optical detections are consistent with both the expectation of an afterglow emission from a structured jet \citep{Rossi2002,Zhang2002,Kathirgamaraju2017,
Lazzati2017a,Lazzati2017b,Gottlieb2017,Gottlieb2018,Salafia2017a,Lazzati2017c} and from the deceleration of an isotropic mildly 
relativistic outflow \citep{Mooley2017,Salafia2017b}.  Both  scenarios can account reasonably well for the low luminosity of the 
prompt emission and the late-time rise up of the X-ray and radio emission, although the slow late-time radio increase might require 
some modifications of the simplest assumptions (\citealt{Mooley2017}; but see \citealt{Lazzati2017c}).

\begin{figure*}[t]
\centering
\includegraphics[scale=0.55]{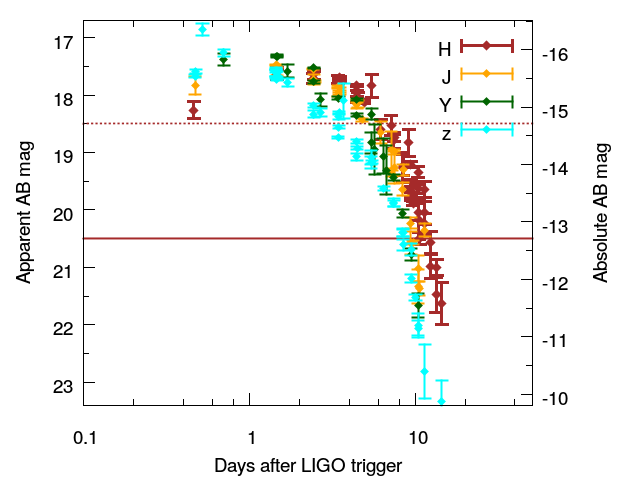}
\caption{
Light curve of the kilonova associated to the gravitational wave/short GRB event GW170817/GRB170817A in the IRT filters 
(from \citealt{Tanvir2017,Pian2017,Arcavi2017,Cowperthwaite2017,Drout2017,Shappee2017,Kasliwal2017,Smartt2017,Troja2017}). 
The continuous and dashed red lines indicate the THESEUS/IRT limiting H magnitudes for imaging and prism 
spectroscopy, respectively, with 300s of exposure \citep[see][]{Amati2017}. Credit: A. Melandri
}
\label{fig:kilonova_data}   
\end{figure*}

In Figure \ref{fig:sens} the measured Fermi/GBM flux in the 50-300 keV of $0.8\pm0.3$ ph cm$^{-2}$ s$^{-1}$ is extrapolated at distances larger 
than the distance of GRB170817A ($\sim$ 40 Mpc) and compared with the THESEUS/XGIS sensitivity in the 30-150 keV band \citep[][see Fig. 36]{Amati2017} 
rescaled to the 50-300 keV. From this plot, we can see that not only this source could had been clearly detected with XGIS but it could 
have been detected up to $\sim 70-80$ Mpc, that is nearly twice the actual distance of the source.  
On the other hand, the faint X-ray emission detected weeks/months after the trigger with a flux of the order of a few times 
$10^{-15}$ erg cm$^{-2}$ s$^{-1}$ \citep{Troja2017}, could not had been detected with the SXI. 
Figure~\ref{fig:kilonova_data} shows how IRT could had clearly detected the NIR counterpart of GW170817 recognised to 
be the expected "kilonova" (or "macronova", see next section).  In particular, the observed emission was bright 
enough during the first 2 days to have allowed IRT low-resolution spectroscopy with 
unprecedented coverage. 

\section{Gravitational wave sources}
\label{gw}

\subsection{NS-NS / NS-BH mergers: Collimated emission from Short GRBs \\} 
\label{collimatedemission}

Compact binary coalescences (CBCs) involving neutron stars (NS) and stellar mass black holes (BH) 
are among the sources of GWs that will be likely detected in spades in the next decade. These systems radiate GWs within 
the most sensitive frequency range of ground-based GW detectors (1-2000~Hz), with large GW energy output, of the order 
of 10$^{-2}$~$M_\odot c^2$, and gravitational waveforms well predicted by General Relativity (see, e.g., \citealt{Baiotti2017} for a review). 
From the merger of two stellar-mass black holes the current consensus is that no significant EM counterpart emission is 
expected, with exceptions where a short GRB-like EM signal may be produced from those BH-BH systems merging in very 
high density environments (as for example in an AGN disk, \citealt{Bartos2016}), or in other exotic conditions 
\citep[e.g.][]{Perna2016,Seto2011,Loeb2016}. 
On the other hand, EM emission is expected on robust theoretical ground for merging NS-NS or NS-BH. 
For more than a decade, mounting indirect evidence supported the long-standing hypothesis that short GRB progenitors 
are associated with CBC systems with at least one neutron star \citep[e.g.][]{Paczynski1986,Eichler1989,Narayan1992,
Barthelmy2005,Fox2005,Gehrels2005}. The most compelling pieces of evidence include (i) the observation of short GRBs 
in both elliptical and late-type star forming galaxies, with a preference for the most massive ones (ii) the relatively 
large projected offsets of these events with respect to the center of their host galaxies, and (iii) the lack of supernova 
associations, as opposed to the case of long GRBs (e.g., \citealt{Berger2014} and refs therein). The most recent 
numerical simulations also provided supporting (though not conclusive) evidence that such merging systems might 
act as short GRB central engines \citep[e.g.][]{Rezzolla2011,Paschalidis2015,Ruiz2016,Kawamura2016,Rosswog2013}. 
Short GRBs have been historically distinguished from long GRBs as those with gamma-ray burst (prompt emission) duration 
less than 2 s \citep{Kouveliotou1993}. However, there are similarities between the two classes of GRBs as far as their prompt 
emission \citep[e.g.][]{Ghirlanda2015b,Ghirlanda2009,Ghirlanda2011} and their afterglows \citep[e.g.][]{D'Avanzo2015}. 
It is likely that the two populations overlaps and contaminate each other, especially when selection is based solely 
on the observed duration (\citealt{Bromberg2013}, but see \citealt{Zhang2012}).  Therefore, GW signals from short GRBs 
might be an additional parameter to be considered to firmly distinguish core-collapse (long) from compact merger (short) 
progenitors. 

Since GW events enable to individuate faint short GRBs as GRB170817A \citep{Abbott2017a,Goldstein2017,Savchenko2017}, 
future joint multi-messenger detections of such sources, ensured by the presence of space missions like THESEUS, 
will shed light on several still debated topics as: 1) the jet intrinsic structure and their properties, and ultimately 
the crucial issue of short GRB energetics; 2) the physics regulating the on- and off-axis emission; 3) the late-time 
(e.g. after 1-2 weeks) component origin. 

\begin{table*}
\caption{Number of NS-NS (BNS) mergers expected to be detected in the next years by second- (2020+) and 
third- (2030+) generation GW detectors and the expected detection number of electromagnetic counterparts as short GRBs 
(collimated) and X-ray isotropic emitting counterparts (see \S 3.1 and 3.2) 
with THESEUS SXI and XGIS (see text for more details). BNS rate is a realistic estimate from \citealt{Abadie2010a} 
and \citealt{Sathyaprakash2012} and the BNS range indicates the sky- and orbital inclination-averaged distance up to 
which GW detectors can detect a BNS with $SNR=8$. 
}
\label{tab:rate}  
\resizebox{\columnwidth}{!}{
\begin{tabular}{llllll}
\hline\noalign{\smallskip}
\multicolumn{3}{c}{GW observations} &  \multicolumn{3}{c}{THESEUS XGIS/SXI joint GW+EM observations} \\
\noalign{\smallskip}\hline\noalign{\smallskip}
Epoch & GW detector & BNS range & BNS rate & XGIS/sGRB rate & SXI/X-ray isotropic \\
& & & (yr$^{-1}$) & (yr$^{-1}$) & counterpart rate  (yr$^{-1}$) \\
\noalign{\smallskip}\hline\noalign{\smallskip}
2020+ & Second-generation (advanced LIGO, & $\sim$200~Mpc & $\sim$40$^*$ & $\sim$5-15 & $\sim$1-3 (simultaneous) \\ 
& Advanced Virgo, India-LIGO, KAGRA) & & & & $\sim$6-12 (+follow-up) \\
2030+ & Second + Third-generation & $\sim$15-20~Gpc & $>$10000 & $\sim$15-35 & $\gtrsim$100 \\
& (e.g. ET, Cosmic Explorer) & & & &  \\
\noalign{\smallskip}\hline
\end{tabular}
}
$^*$ from Abadie et al. 2010a
\end{table*}
 
Table \ref{tab:rate} shows the expected rate of THESEUS/XGIS short GRB detections with a GW counterpart from merging NS-NS systems (i.e. within 
the GW detector horizon). The quoted numbers are obtained by assuming that all on-axis short GRBs will be detected with THESEUS/XGIS within the distance range of the 2G GW detectors (see Figure \ref{fig:rate}). 
We correct the realistic estimate of NS-NS GW detection rate, $\sim40$ yr$^{-1}$ 
(\citealt{Abadie2010a}, see also \citealt{Belczynski2017}) for the fraction of the sky covered by the XGIS FoV, that is $\sim$50\%, and 
the short GRB jet collimation factor by assuming a jet half-opening angle range of 10--40 deg. We also considered the possibility to 
observe off-axis short GRBs up to 5 times a jet half-opening angle of 10 degrees and 2 times a jet of 40 
degrees \citep{Kathirgamaraju2017,Pescalli2016}. The 5-times factor was obtained by considering 
the THESEUS/XGIS 1 sec photon flux sensitivity $\sim$0.2 ph cm$^{-2}$ s$^{-1}$ (see also \citealt{Amati2017}).  
We are here assuming that every BNS merger produces a jetted short GRB, which is still an open issue. 
Results show that, during the 2020's, the GW+EM detection rate of short GRBs  with THESEUS is found to be of the order of 5--15 per year. 
By the time of the launch of THESEUS, gravitational radiation from NS-NS and NS-BH mergers will be detectable by third-generation detectors 
such as the Einstein Telescope (ET) up to redshifts z$\sim$2 or larger \citep[see, e.g.,][]{Sathyaprakash2012, Punturo2010}, thus dramatically 
increasing the GW+EM on-axis short GRB detection rate. 
The important implication is that almost all THESEUS short GRBs will have a detectable GW emission. Indeed, it is likely that at the typical 
distances at which ET detects GW events, the only EM counterparts that could feasibly be detected are short GRBs and their afterglows, 
making the role of THESEUS crucial for multi-messenger astronomy by that time.

\subsection{NS-NS / NS-BH mergers: non-collimated soft X-ray and optical/NIR emission} 
\label{isotropicemission}

In this section we introduce another possible electromagnetic counterpart  of CBC events that is not collimated and is expected to peak in the soft X-ray band. 
GW emission from CBCs depends only weakly on the inclination angle of the inspiral orbit and therefore these events are in general observable 
at any viewing angle. As a consequence, most of the GW-detected mergers are expected to be observed off-axis (i.e. with a large angular 
distance of the observer from the orbital axis). This makes the non collimated, nearly isotropic EM components extremely relevant for 
the multi-messenger investigation of CBCs. 

A potentially powerful nearly-isotropic emission is expected if a NS-NS merger produces a long-lived millisecond magnetar. 
In this case, soft X-ray to optical transients can be powered by the magnetar spin-down emission reprocessed by the baryon-polluted 
environment surrounding the merger site (mostly due to isotropic matter ejection in the early post-merger phase), with time scales of minutes 
to days and luminosities in the range 10$^{43}$--10$^{48}$~erg~s$^{-1}$ \citep[e.g.][]{Yu2013,Metzger2014,Siegel2016a,Siegel2016b}. 
In particular, in soft X-rays (at $\sim$keV photon energies) these transients can last from minutes to hours and, for the most optimistic models, 
reach luminosities as high as 10$^{48}$~erg~s$^{-1}$ \citep{Siegel2016a,Siegel2016b}. 
According to alternative models, X-ray emission may also be generated via direct dissipation of magnetar winds \citep[see, e.g.,][]{Zhang2013, Rezzolla2015}. 
Furthermore, the high pressure of the magnetar wind can in some cases accelerate the expansion of previously ejected matter into the interstellar 
medium up to relativistic velocities, causing a front shock which in turn produces synchrotron radiation in the X-ray band 
(with a high beaming factor of $\sim$0.8; see, e.g., \citealt{Gao2013}).

\begin{figure*}[t!]
\centering
\includegraphics[scale=0.23]{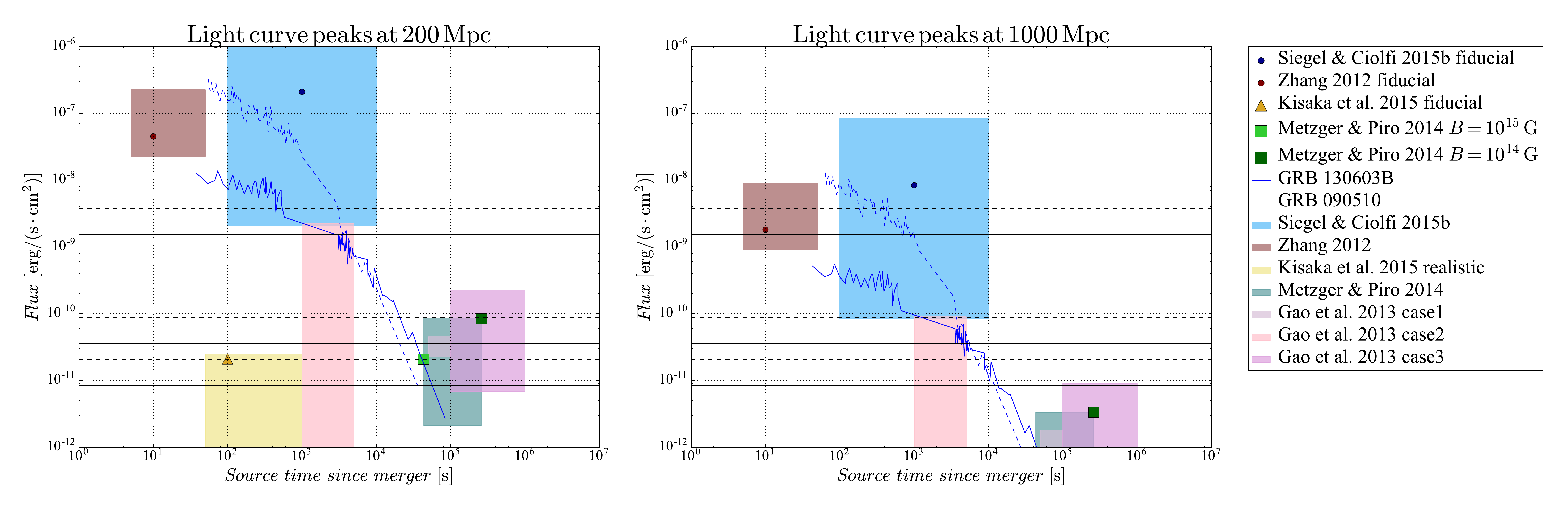}
\caption{Expected X-ray fluxes at peak luminosity from two different luminosity
distances (z=0.05 on the left panel, and z=0.2 on the right panel) and
from different models of magnetar-powered X-ray emission from long-lived
NS-NS merger remnants. Predictions from each model are represented by a
coloured region and/or by single dots that are indicative of fiducial
cases (see the legend on the right).
The blue lines show two short GRB X-ray afterglows observed with
Swift/XRT.
The black curves show the SXI sensitivity from higher to lower values at
10s, 100s, 1ks and 10ks, assuming a source column density of $5\times10^{20}$ 
cm$^{-2}$ (i.e., well out of the Galactic plane and very little intrinsic
absorption, solid lines) and $10^{22}$ cm$^{-2}$ (significant intrinsic
absorption, dashed lines).} 
\label{fig:X-rayFlux}   
\end{figure*}

Figure~\ref{fig:X-rayFlux} shows predictions for magnetar-powered X-ray emission following a NS-NS merger according to a number of different models. 
Overall, typical time scales for these transients are comparable to magnetar spin-down time scales of $\sim$10$^3$--10$^5$~s and the predicted 
luminosities span a wide range that goes from~10$^{41}$ to ~10$^{48}$~erg~s$^{-1}$. 
Joint GW+EM detection rates with THESEUS/SXI are discussed below. 
These rates depend not only on the rate of NS-NS mergers, but also on the (essentially unknown) fraction of mergers forming a long-lived NS remnant, 
which is necessary to produce spindown-powered transients. 
The observation of this type of emission after a NS-NS merger would indeed indicate that the remnant is long-lived, allowing for significant 
constraints on the equation of state of the remnant itself \citep[e.g.][]{Piro2017,Drago2018}.

In the case of GW170817/GRB170817A, no evidence for this type of emission was found in the soft X-ray band. However, the first deep pointed 
observations at $\sim$keV photon energies only started $\sim$15 hours after merger with Swift/XRT \citep{Evans2017}. Possible constraints 
could be provided by the MAXI (2-10 keV) observations taken at 4.5 hours after the trigger, with a flux limit of $\sim10^{-8}$ erg cm$^{-2}$ s$^{-1}$.  
We note that for GW170817 the nature of the remnant (BH vs. long-lived NS) was not established for this event, thus making it difficult to put constraints 
on theoretical expectations.  
For future observations, being able to catch the soft X-ray emission (or to firmly assess its absence) within the relevant time scale after a 
GW trigger will require a monitoring (wide-field) instrument sensitive to $\sim$keV energies. THESEUS/SXI will perfectly respond to this need.

The expected detection rate of the isotropic X-ray emission from NS-NS mergers is quoted in Table \ref{tab:rate}. Taking into account the sensitivity vs exposure time provided in \citealt{Amati2017} the SXI will detect almost all X-ray transients at $<200$ Mpc, as shown in Figure \ref{fig:X-rayFlux}.  Starting from the realistic rate of NS-NS mergers that will be detected with GW observatories in 2020-2030, we have accounted for: 1) the fraction of the sky covered by the SXI FoV, that is $\sim8$\%, for serendipitous discoveries, and 2) the fraction of NS-NS systems that can produce X-ray emission (i.e. that do not form immediately a BH), that we assumed to be within 30\%-60\% \citep{Gao2013,Piro2017}. Moreover, we consider the fraction 
of BNS sources that could be followed-up with SXI after a GW alert, estimated to be of the order of $\sim40$\%.  From these computations, we find that during the 2020s the joint GW+EM detection rate with THESEUS of these X-ray counterparts of 
NS-NS mergers is $\sim$ 6-12 per year. Starting from around 2030, with the third-generation GW detectors, isotropic X-ray emission from NS-NS mergers 
as predicted by some models \citep[e.g.][]{Siegel2016b} could be detected up to $\sim10$ times larger distances, with an improved 
joint GW+EM detection rate of few hundreds per year (depending on the largely uncertain intrinsic luminosity of such X-ray component, 
see Table~\ref{tab:rate} and Fig. \ref{fig:X-rayFlux}). With such statistics, THESEUS will provide a unique contribution to characterise 
this X-ray emission from NS-NS systems. 

Another well known type of nearly-isotropic emission expected from CBCs involving NSs is the so-called ``kilonova'' or 
``macronova'' \citep[e.g.,][]{Li_Paczynski1998,Metzger2010}. 
NS-NS or NS-BH mergers can eject a substantial amount of matter (up $10^{-2}~M_\odot$ or more) which becomes unbound and leaves the system. 
This material can be expelled both during merger (dynamical ejecta) and in the post-merger phase, in the form of baryon-loaded winds from 
the accretion disk surrounding the merger remnant (or from the remnant NS itself, for NS-NS mergers without prompt collapse to BH). 
Due to the unique conditions of high neutron density and temperature, r-process nucleosynthesis of very heavy elements takes place in 
the ejected matter and days after merger the radioactive decay of such elements heats up the material producing a thermal transient signal 
peaking in the optical/NIR bands and with typical luminosities of $\sim$10$^{40}-$10$^{41}$~erg~s$^{-1}$ \citep[see, e.g.,][]{Fernandez2016,Metzger2017}.
The temporal and spectral properties of these signals encode crucial information on the nature of the merger progenitor (e.g., NS-NS or NS-BH), 
the equation of state of neutron stars, and the heavy element chemical enrichment of the Universe.

Before the GW170817 event, observational and photometric-only evidence of kilonova/macronova transients relied only on a few candidates 
observed during short GRB follow-up campaigns \citep[e.g.][]{Tanvir2013,Jin2015,Berger2013}. 
The recent discovery of AT2017gfo has now provided the first 
compelling evidence, both photometric and spectroscopic, of the existence of kilonovae/macronovae (\citealt{Pian2017}, see also 
\citealt{Abbott2017b,Tanvir2017,Nicholl2017,Smartt2017,Tanaka2017,Chornock2017} ). 
The observations of AT2017gfo consolidated the presence of a strong IR emission component (reaching its maximum 1.5 days after 
merger, with 17.2 and 17.5 mag in the J and K bands, respectively \citep[e.g.][]{Tanvir2017}.
This provides a very strong science case for the IR instrument on-board THESEUS (IRT) as shown in Figure~\ref{fig:kilonova}.  

Summarizing, both serendipitous discoveries within the large THESEUS/SXI FoV and re-pointing of THESEUS in response to a GW trigger will 
allow to study off-axis X-ray and NIR emission expected from NS-NS systems. With THESEUS/SXI in 
combination with the second-generation detector network, almost all predicted non-collimated X-ray counterparts of GW events 
from NS-NS merging systems will be easily detected simultaneously with the GW trigger and/or with rapid follow-up of the 
GW-individuated sky region.  Among the open questions that THESEUS will help to address there are: 1) does the NS-NS merger 
create a NS or a BH, and how fast?; 2) how much matter is expelled in the NS mergers? At which speeds?; 3) what is the amount 
of asymmetry in the NS-NS merger ejecta and the corresponding optical emission?

\subsection{Core-collapse of massive stars: supernovae and long GRBs \\} 

Core-collapse supernovae (CCSNe) represent another type of GW sources that are of great interest for the involved community. 
However, contrary to the CBC case, their expected GW emission is highly uncertain as it strongly depends on the rather unknown SN explosion mechanism \citep[e.g.][]{Logue2012,Powell2016}. 
Not only the signal morphology (waveform), but also the expected energy output are still under debate.  
Thus, depending on the assumed model as well as dedicated data analysis techniques, during the second-generation GW detector 
network era the  detection of GW from CCSNe is predicted up to few Mpc or less \citep[e.g.][]{Ott2010}, or up to $\sim$100 Mpc \citep[e.g.][]{VanPutten2017}. 
While this makes it difficult to predict the GW signal and its detectability, it also represents a unique opportunity 
to probe the CCSN inner dynamics that cannot be explored via the sole observation of EM signals. 

The firm association of nearby long GRBs with temporally and spatially coincident CCSNe 
\citep[e.g.][]{Woosley2006,Galama1998,Stanek2003}\footnote{For GRB 060614 \citep{DellaValle2006,Fynbo2006}, in spite of marginal evidence 
for an associated kilonova, which would make it a short GRB, this event, along with GRB 060505, leaves the possibility of long 
SN-less GRBs, see also \citealt{Xu2009}. } implies that any long GRB, if close enough, should be associated with a detectable 
GW emission and thus offers a very interesting potential synergy between gamma-ray and GW detectors. 
The current uncertainties on the GW detection horizon of these events imply rates that 
can be as low as a few events per century, but third generation GW detectors such as the Einstein Telescope will offer much better prospects with possible horizons up to 1 Gpc.  The first joined GW/GRB/SN observations, possibly combined also with neutrino detections (Sect. \ref{nu}), will prove crucial 
to unravel the nature of these sources and their explosion mechanism. Observed long GRB rate density is $\sim 1$ Gpc$^{-3}$ yr$^{-1}$  \citep[e.g.][]{Le2007} thus simultaneous GW+EM detection 
rate of possibly more than 1 event per year is realistically expected not before the third-generation GW interferometers. 
Off-axis X-ray afterglow detections (``orphan afterglows'') \citep[e.g.][]{Granot2002,Ghirlanda2013,Ghirlanda2015a} can 
potentially increase the simultaneous GW+EM detection rate for nearby long GRBs by a factor that strongly depends on the 
jet opening angle and the observer viewing angle. 
THESEUS may also observe the appearance of a NIR orphan afterglow few days after the reception of a GW signal due to a 
collapsing massive star. In addition, the possible large number of low luminosity GRBs \citep[LLGRBs, e.g. ][]{Toma2007,Virgili2009} 
in the nearby Universe, expected to be up to 1000 times more numerous than long GRBs, will provide clear signatures in the 
GW detectors because of their much smaller distances with respect to long GRBs. 

Beside the GRB-connected phenomenology, Wolf-Rayet stars as well as red and blue supergiants are expected 
to exhibit bright shock breakout soon after their core collapse, 
with X-ray bursts lasting 10-1000 s and with luminosities expected in the range $10^{43}-10^{46}$ erg s$^{-1}$. These progenitors 
are likely responsible for Type Ibc and most Type II SNe, which occur at rates of $2.6\times10^{-5}$ and $4.5\times10^{-5}$ Mpc$^{-3}$ yr$^{-1}$, 
respectively \citep{Li2011}. 
THESEUS/SXI and XGIS can detect these events up to $\sim50$ Mpc leading to a rate of a few per year \citep{Amati2017}. 
We expect up to few shock breakout events per year that can be detected with THESEUS/SXI simultaneously with their GW counterpart 
during the 3G GW detector era.  
Shock Breakout (SBO) components are temporally closer to the possibly associated GW events than the optical CCSNe counterpart, 
thus their detection can mark with more precision the start time of the gravitational radiation emission and can be 
used in the challenging signal search processes \citep{Andreoni2016}.

\subsection{Magnetars \\} 
Fractures of the solid crust on the surface of highly magnetised neutrons stars and/or dramatic magnetic field readjustments represent 
the most widely accepted explanations to interpret the magnetar bursting activity and in particular the rare giant flares 
observed in X-rays from three different soft gamma repeaters (SGRs; see, e.g., \citealt{Thompson1995,Guidorzi2004,Mereghetti2015}).
The initial short ($<0.5$s) bright spikes of SGR can be detected with the XGIS to considerable distances. The XGIS low energy threshold is better suited for the detection of such events with respect to other coded-mask detectotrs. Flares and bursting episodes will be easily detectable with SXI  \citep{Amati2017}.

The above events will inevitably excite non-radial oscillation modes that may produce detectable GWs 
\citep[see, e.g.,][]{Corsi2011, Ciolfi2011}. The most recent estimates for the energy reservoir available in a 
giant flare are between 10$^{45}$~erg (about the same as the total EM emission) and 10$^{47}$~erg. The efficiency of 
conversion of this energy into GWs was estimated in numerical relativity simulations and it was found to be likely 
too small to be within the sensitivity range of present GW detectors \citep{Ciolfi2012, Lasky2012}. However, at the 
typical dominant (i.e. f-mode) oscillation frequencies in NSs ($\sim$kHz), ET will be sensitive to much lower GW 
energies \citep{Punturo2010}. 
Therefore, a relatively close giant flare event might lead to a detectable GW emission.

\section{Neutrino sources \\}                      
\label{nu}            

Several gamma-ray and X-ray sources that THESEUS will observe as GRBs, CCSNe and AGNs, are also expected to originate neutrinos. 
Due to their low interaction cross-section, neutrinos can probe the innermost regions similarly to gravitational waves but, 
in addition, neutrino detectors can provide a more refined sky localisation than GW interferometers, with an uncertainty that 
goes from few degrees down to a fraction of a degree. Current neutrino deep-water-based detectors include DUMAND, Lake Baikal, 
and ANTARES. These Northern hemisphere detectors  complement the South Pole based IceCube, the first km-scale neutrino observatory, 
completed and in full operation since 2010. Two major upgrades for the near and far future are planned with the construction of 
Km3Net in the Northern hemisphere, started in 2015, and IceCubeGen2, an upgrade to a 10 km$^3$ detector of IceCube 
\citep[e.g.][and references therein]{Aartsen2014}. These are prevalently high-energy neutrino detectors but IceCube 
and Km3Net can detect also MeV neutrinos due to the capabilities to suppress background rate, together with other liquid 
scintillators and liquid Argon Time-Projection Chamber detectors \citep[e.g. see reviews by][and references therein]{Scholberg2012,Gil-Botella2016}. 

Pulses of low energy neutrinos ($<$10~MeV) are expected to be released during CCSNe with an energy release up to 10$^{53}$~erg. 
Indeed, MeV neutrinos have been detected so far only from one CCSN, namely SN1987A, in the Large Magellanic Cloud at 50~kpc 
distance \citep[e.g.][]{Gaisser1995}. Comparison of the SN1987A neutrino signal with theoretical predictions showed that the 
general features of CCSNe are compatible with the observations \citep[e.g.][]{Giunti2007}. However, significant uncertainties 
are still affecting cCCSNe modelings and, more in general, the core-collapse processes of massive stars. Great advances are 
expected from GW and further neutrino detections that will be achieved with the next generation detectors.  Both GW and 
neutrinos can provide important information as the degree of asymmetry in the matter distribution, as well as the rotation 
rate and the strength of the magnetic fields, that can be used as priors in numerical simulations \citep[see, e.g.,][and reference therein]{Chassande-Mottin2010}. 

Significant evidence of high-energy (TeV-PeV) cosmic neutrinos has recently been obtained from an extensive IceCube fourth-year 
data analysis \citep[e.g.][]{Aartsen2014b}. The lack of significant anisotropy in the data sky direction distribution is consistent 
with an (at least partially) extragalactic origin of the neutrino sources.  High-energy neutrinos originate from hadrons acceleration, 
for example in jets, where after interacting with high energy photons produce charged pions decaying as high energy neutrinos 
\citep[$>$10$^5$~GeV; see, e.g.,][]{Waxman1997}.  Among the candidates that have been proposed to be responsible for the observed 
high-energy neutrino flux there are GRBs, AGN and blazars that are part of the main THESEUS targets in the context of the 
Time-domain Universe \citep{Amati2017}. THESEUS/SXI will enable to monitor the X-ray flux of hundreds of AGN on daily timescales and provide an unprecedented look at long-term variability of large samples of AGN and Blazars.  Deep monitoring will also performed with XGIS in the hard X-ray spectral regime. These observations will provide an ideal tool for neutrino simultaneous detection. 

GRB are historically addressed among the best candidates of Ultra High Energy Cosmic Rays (UHCR) \citep[e.g.][]{Ghisellini2008}, 
together with AGNs. Recent results from the Pierre Auger Observatory found evidence for dipolar anisotropy in CR at $E>8\times10^{18}$ eV 
towards a given direction in the sky, which is compatible with an extra-galactic origin, with possible suggestion that they are 
due to Large Scale Structures, with relatively nearby sources within 300 Mpc \citep[e.g.][]{Globus2017}. As UHECR sources, GRBs 
are therefore  addressed as promising high-energy neutrinos source candidates together with AGN. However, searches for neutrino 
events in coincidence with GRBs have not provided any confirmed association so far, possibly because of the average large distances 
of GRBs and/or a low neutrino production efficiency in bright GRBs.  
On the other hand, GRBs sources are particularly interesting since they could potentially emit also GWs. Possible detection could be 
achieved with the next generation of neutrino detectors. 
For long GRB, neutrinos emitted along the jet direction give the highest chances of detection. The expected rate of on-axis GRB that 
can be detected with IceCube has been estimated to be of the order of $\sim$0.3 per year 
\citep[e.g.][]{Xiao2017}.
The lack of neutrinos from the very nearby short GRB associated with the GW170817 source has been interpreted 
to be due to the off-axis viewing angle of our line of sight with respect to the jet direction\citep{Albert2017}, but the feasibility of future joint EM and GW/neutrino observations are supported by theoretical background. In particular, for the 
case of short GRBs, according to the most recent studies \citep{Kimura2017}, high-energy neutrinos are thought to be most 
efficiently produced during the so called "Extended Emission" (EE), a softer, prolonged emission lasting few tens up to hundreds 
of seconds, that follows the initial count rate spike that characterises some short GRBs \citep{Norris2006,Kaneko2015}.
It has been observed that about $\sim25$\% of short GRBs are accompanied by an EE \citep{Sakamoto2011}. This fraction is likely 
biased by the lack of X-ray survey instruments that could detect this component and likely more short GRBs are accompanied by 
EE \citep{Nakamura2014}, possibly up to 50\% \citep{Kimura2017}.
According to the neutrino detection probability estimates as a function of the short GRB with EE distance computed by \citet{Kimura2017}, we expect that 
THESEUS/neutrino counterpart detection rate of on-axis short GRBs with EE within the horizon of IceCube and IceCubeGen2 is of the order 
of 0.02-0.25 and 0.1-0.5 per year, respectively. 
By considering the possibility to observe neutrinos also from short GRB with EE viewed off-axis, the THESEUS/neutrino counterpart 
detection rate may increase up to 0.2-4, and 0.5-7 per year, respectively. 
Future multi-messenger campaigns with deeper detector sensitivities will likely further constrain GRB progenitor models, 
clarifying the presence of a jet and its composition, and the relative neutrino/EM energy budgets and the role 
of GRBs as sources of UHCRs  \citep{Abbasi2012}. 

THESEUS/SXI and XGIS can detect SN shock breakouts events up to $\sim50$ Mpc (see previous section), thus leading to a potential joint neutrino 
detections of a few events per year with new generation  neutrino detectors as Km3Net or IceCubeGen2. 
Blazars have been considered among the possible source candidates for the recently detected IceCube cosmic high-energy neutrino flux. 
THESEUS/blazars detection rate is estimated to be of hundreds per year (see Tab.2 in Amati et al. 2017).

\section{Summary}

The first detection of the electromagnetic counterparts of a GW source has confirmed a number of theoretical expectations and 
boosted the nascent multi-messenger astronomy. In this review we have discussed several classes of sources, including compact 
binary coalescences, core-collapsing massive stars, and instability episodes on NSs that are expected to originate simultaneously 
high-frequency GWs, neutrinos and EM emission across the entire EM spectrum, including in particular high energy emission 
(in X-rays and gamma-rays). We have shown that the mission concept THESEUS has the potential to play a crucial role in the 
multi-messenger investigation of these sources. 
THESEUS, if approved, will have the capability to detect a very large number of transient sources in the X-ray and gamma-ray 
sky due to its wide field of view, and to automatically follow-up any high energy detection in the near infrared. 
In addition, it will be able to localise the sources down to arcminute (in gamma and X-rays) or to arcsecond (in NIR). 
The instrumental characteristics of THESEUS are ideal to operate in synergy with 
the facilities that will be available by the time of the mission: several new generation ground- and space-based 
telescopes, second- and third-generation GW detector networks and 10 km$^3$ neutrino detectors. This makes THESEUS 
perfectly suited for the coming golden era of multi-messenger astronomy and astrophysics.

\section*{Acknowledgements} 
{\bf Maria Giovanna Dainotti acknowledges funding from the European Union through the Marie Curie Action FP7-PEOPLE-2013-IOF, under grant agreement No. 626267 ("Cosmological Candles"). Sergio Colafrancesco is supported by the South African Research Chairs Initiative of the Department of Science and Technology
and National Research Foundation of South Africa (Grant no 77948).}

\bibliographystyle{model1a-num-names}
\bibliography{theseus_mma}

\end{document}